\documentclass[a4paper,10pt,aps,pre,twocolumn,amsmath,amssymb]{revtex4}

\usepackage{graphicx,color,doi}

\hypersetup{colorlinks,linkcolor=blue,urlcolor=red,citecolor=blue}

\newcommand{\ahum}[1]{``#1''}
\newcommand{\eq}[1]{Eq.~(\ref{#1})}
\newcommand{\fig}[1]{Fig.~\ref{#1}}

\newcommand{\olcite}[1]{Ref.~\onlinecite{#1}}
\newcommand{\mean}[1]{\langle #1 \rangle}

\newcommand{\pc}{p_{\rm t}}
\newcommand{\dd}{{\rm d}}
\newcommand{\rmin}{\rho_{\rm min}}
\newcommand{\rmax}{\rho_{\rm max}}

\newcommand{\pg}{p_{\chi}(L)}
\newcommand{\km}{{k_{\rm max}}}
\newcommand{\xm}{\chi_{\rm max}(L)}
\newcommand{\mustar}{\mu^\star_L}

\begin{document}

\title{Crossover from a Kosterlitz-Thouless to a discontinuous phase transition 
in two-dimensional liquid crystals}

\author{Richard L. C. Vink}

\affiliation{Institute of Theoretical Physics, Georg-August-Universit\"at 
G\"ottingen, Friedrich-Hund-Platz~1, D-37077 G\"ottingen, Germany}

\date{\today}

\begin{abstract} Liquid crystals in two dimensions do not support long-ranged 
nematic order, but a quasi-nematic phase where the orientational correlations 
decay algebraically is possible. The transition from the isotropic to the 
quasi-nematic phase can be continuous of the Kosterlitz-Thouless type, or it can 
be first-order. We report here on a liquid crystal model where the nature of the 
isotropic to quasi-nematic transition can be tuned via a single parameter $p$ in 
the pair potential. For $p<\pc$, the transition is of the Kosterlitz-Thouless 
type, while for $p>\pc$ it is first-order. Precisely at $p=\pc$, there is a 
tricritical point, where, in addition to the orientational correlations, also 
the positional correlations decay algebraically. The tricritical behavior is 
analyzed in detail, including an accurate estimate of $\pc$. The results follow 
from extensive Monte Carlo simulations combined with a finite-size scaling 
analysis. Paramount in the analysis is a scheme to facilitate the extrapolation 
of simulation data in parameters that are not necessarily field variables (in 
this case the parameter $p$) the details of which are also provided. This scheme 
provides a simple and powerful alternative for situations where standard 
histogram reweighting cannot be applied. \end{abstract}

\maketitle

\section{Introduction}

Anisotropic molecules confined at plates~\cite{citeulike:4006159, 
citeulike:2811025, citeulike:3991276} or interfaces~\cite{Jordens2013} give rise 
to liquid crystalline systems that are effectively two dimensional. 
Consequently, there is much interest to understand the nature of the order 
(isotropic, nematic) that arises, and the associated phase transitions. For 
two-dimensional (2D) liquid crystals, the accepted view is that long-ranged 
nematic order does not exist in the thermodynamic limit~\cite{physreva.4.675}. 
There is, however, the possibility of quasi-nematic order, whereby the nematic 
order decays algebraically with distance. Computer simulations of 2D rods and 
needles indeed reveal that quasi-nematic order arises, provided the particle 
density is high enough~\cite{physreva.31.1776, bates.frenkel:2000, 
citeulike:5931786}, and this order persists even in slit-pores having a finite 
width~\cite{lagomarsino.dogterom.ea:2003}. The transition from the isotropic 
phase, where nematic order decays exponentially, to the quasi-nematic phase is 
continuous in these systems, and of the Kosterlitz-Thouless (KT) 
type~\cite{kosterlitz.thouless:1973, kosterlitz:1974}.

While the existence of a KT transition in 2D rods and needles is thus well 
established by these simulations, conclusive experimental evidence for such a 
(continuous) transition remains difficult to obtain~\cite{citeulike:4067023}. 
Typically, experiments reveal pronounced two-phase 
coexistence~\cite{citeulike:4006159, citeulike:2811025, citeulike:3991276}, 
suggesting that the isotropic $\leftrightarrow$ quasi-nematic transition is 
first-order, which is at variance with the 
conventional~\cite{kosterlitz.thouless:1973, kosterlitz:1974} KT scenario. A 
possible explanation is provided by van Enter and Shlosman, who rigorously 
proved that the KT transition can also become first-order, provided a certain 
condition in the pair potential is met~\cite{physrevlett.89.285702, 
enter.shlosman:2005, citeulike:12935130}. Inspired by this proof, Wensink and 
Vink proposed a liquid crystal model in which a first-order isotropic 
$\leftrightarrow$ quasi-nematic transition could indeed be 
realized~\cite{vink.wensink:2007}. The order parameter of this transition is the 
density, which is low (high) in the isotropic (quasi-nematic) phase, and so 
there is a density gap. At the transition, which can be driven by varying the 
chemical potential, the density \ahum{jumps} discontinuously between the low and 
high value, as is characteristic of a first-order transition. In addition, at 
the transition, simulation snapshots reveal pronounced coexistence between 
isotropic and quasi-nematic domains, furthermore confirming that the transition 
is first-order.

The isotropic $\leftrightarrow$ quasi-nematic transition in 2D liquid crystals 
can thus manifest itself in two forms, namely as (1) a continuous KT transition, 
or (2) a first-order transition. This suggests the possibility of tricritical 
behavior in these systems, where the transition type changes from first-order to 
continuous~\cite{Lawrie1984}. The purpose of this paper is to show that a 
tricritical point can indeed be identified. At the tricritical point, in 
addition to the orientational correlations, also the density correlations become 
quasi-long-ranged, i.e.~the radial distribution function $g(r)$ asymptotically 
decays as a power law. In contrast, everywhere else in the phase diagram, $g(r)$ 
is short-ranged, decaying exponentially. Our results follow from Monte Carlo 
simulations combined with a finite-size scaling analysis. Of particular note is 
the use of a new extrapolation scheme, similar in spirit to histogram 
reweighting~\cite{ferrenberg.swendsen:1988}, but one which can also be applied 
to variables that are not necessarily field variables. The use of this scheme 
greatly reduces the computational cost of the simulations.

\section{Model and methods}

\subsection{2D liquid crystal model}

We use the liquid crystal model of \olcite{vink.wensink:2007} whose pair 
potential is strictly short-ranged and given by
\begin{equation}\label{eq:vw}
 E = \sum_{i=1}^N \sum_{j=i+1}^N \epsilon \left(
 1 - |\vec{d}_i \cdot \vec{d}_j|^p \right) H(a-r_{ij}) \quad,
\end{equation}
with $N$ the number of particles, $r_{ij}$ the distance between (point) 
particles $i$ and $j$, interaction range $a$, $H(x)$ the Heaviside unit step 
function, and $\epsilon$ a coupling constant to set the temperature scale (in 
what follows, $a$ is the unit of length, $\epsilon/k_BT=2.5$, with $k_B$ the 
Boltzmann constant). The particle positions are confined to the 2D 
plane; the particle orientations are encoded by the vectors $\vec{d}_i$, taken 
to be 2D unit vectors. In \eq{eq:vw}, a pair of particles $i$ and 
$j$ within a distance $a$ can lower the energy by aligning, either in parallel 
or anti-parallel directions (the absolute value $|\cdot|$ ensures that the 
system is invariant under inversion of the particle orientation, as is 
appropriate for liquid crystals). 

The parameter $p$, which is a positive real number, sets the sharpness of the 
interaction. As $p$ gets larger, the potential becomes increasingly selective 
about the degree of alignment. In the limit $p \to \infty$, a pair of particles 
$i$ and $j$ would lower the energy only when the alignment of the vectors 
$\vec{d}_i$ and $\vec{d}_j$ is perfect. As was shown by van Enter and 
Shlosman~\cite{physrevlett.89.285702, enter.shlosman:2005, citeulike:12935130}, 
a sufficiently large value of the sharpness parameter $p$ is what gives rise to 
first-order phase transitions in these systems. For the model of \eq{eq:vw}, the 
existence of a first-order phase transition for large $p$ was confirmed in 
\olcite{vink.wensink:2007}.

\subsection{Grand canonical Monte Carlo}

We performed grand canonical Monte Carlo simulations of \eq{eq:vw}, i.e.~at 
fixed chemical potential $\mu$, and fluctuating particle number $N$ (simulation 
cells are $L \times L$ squares with periodic boundaries). We used standard 
single particle insertion and deletion moves, each attempted with equal {\it a 
priori} probability, and accepted conform the Metropolis 
criterion~\cite{frenkel.smit:2001}. The principal output of the simulations is 
the distribution $P(N)$, which is the probability of observing a state 
containing $N$ particles. To ensure $P(N)$ is accurately measured, the 
simulations used a biased potential, $V_{\rm sim}=E+f(N)$, $E$ given by 
\eq{eq:vw}, and $f(N)$ a bias function constructed to achieve uniform sampling 
in $N$. An initial estimate of $f(N)$ was obtained using 
\olcite{citeulike:3577799}, in which Wang-Landau 
sampling~\cite{wang.landau:2001} and transition matrix 
sampling~\cite{citeulike:12476499} are combined. The transition matrix elements 
were computed for zero chemical potential~\cite{citeulike:3948984} from which 
$P(N|\mu=0)$ can be constructed. The latter is readily extrapolated to a 
different chemical potential $\mu_1$ via histogram 
reweighting~\cite{ferrenberg.swendsen:1988}
\begin{equation}\label{eq:mu}
 P(N|\mu=\mu_1) \propto P(N|\mu=0) \, e^{\mu_1 N / k_BT} \quad.
\end{equation}
For \eq{eq:vw}, the relevant density range $\rmin=1.35 \leq \rho=N/L^2 \leq 
\rmax=3.5$~\cite{vink.wensink:2007}, to which our simulations were restricted. 
For large $L$, we \ahum{parallelized} by dividing the range into $\sim 10$ 
intervals, and assigning a single processor to each interval. Since the 
transition matrix elements are all collected for the same chemical potential 
($\mu=0$), the matrix elements obtained for each interval may simply be added 
afterward.

\subsection{An alternative to histogram reweighting}

The distribution $P(N)$ depends on all the model parameters, in particular the 
sharpness parameter $p$, the chemical potential $\mu$, and the system size $L$. 
To accurately locate phase transitions requires data for several $L$, such that 
a finite-size scaling analysis can be performed. In addition, we require data 
over a fine range in $p$. This poses a challenge because $p$ is {\it not} a 
field variable, i.e.~it cannot be expressed as a prefactor of some term in the 
Hamiltonian (unlike $\epsilon$, which is a prefactor of $E$, or the chemical 
potential, which induces a term $\mu \times N$). Rather, by changing $p$, the 
{\it shape} of the potential is altered, and hence the underlying density of 
states. Consequently, there is no histogram reweighting analogue of \eq{eq:mu} 
for $p$, and extrapolations in the latter will require a radically different 
approach.

To this end, we note that $P(N)$ is just the canonical partition sum, $P(N) = 
{\rm Tr}_N \{e^{-E}\}$, $E$ given by \eq{eq:vw}, with the trace over the 
positions and internal degrees of freedom of $N$ particles. Hence, $\dd\ln 
P(N)/\dd p= \mean{X_1}_N$, the latter being the canonical expectation value of 
\begin{equation}
 X_a \equiv \sum_{[ij]} \epsilon \, y_{ij}^p \, (\ln y_{ij})^a, \quad 
 y_{ij} = |\vec{d}_i \cdot \vec{d}_j| \quad,
\end{equation}
where the sum is over all pairs for which $r_{ij}<a$. Similarly, $\dd^2\ln 
P(N)/\dd p^2 = \mean{X_2+X_1^2}_N - \mean{X_1}_N^2$. The canonical averages 
$\mean{\cdot}_N$ are trivially collected in a grand canonical simulation: At the 
end of each attempted move, one simply \ahum{updates} the average of $X_1$ and 
$X_2+X_1^2$ for the current number of particles. This requires very little extra 
memory (only two additional arrays are needed) and the CPU cost is also 
negligible, since most quantities needed to compute $X_a$ are already needed for 
the energy calculation (by using a link-cell list, the computational effort per 
Monte Carlo move remains independent of $N$). The extrapolation of $P(N|p_1)$ 
measured at sharpness parameter $p_1$ to a different value $p_2$ then becomes a 
Taylor expansion
\begin{equation}
\label{eq:vink}
\begin{split}
 \ln P(N|p_2) \approx \ln P(N|p_1) + \mean{X_1}_N \Delta p \\ + 
 \frac{1}{2} \left( \mean{X_2+X_1^2}_N - \mean{X_1}_N^2 \right) 
 \Delta p^2 \quad,
\end{split}
\end{equation}
$\Delta p = p_2-p_1$, with the canonical averages obtained at $p_1$ (higher 
order terms can optionally be included, but become increasingly cumbersome to 
calculate; our second-order scheme works well in practice, it can reliably 
extrapolate over a range $\Delta p = \pm 2.5$ or so).

To facilitate finite-size scaling, $L=10-40$ was considered. For each $L$, $\km 
\sim 15$ different values of the sharpness parameter $15 < p_k < 75$ were 
simulated, distributed evenly over the range of interest ($k=1,\ldots,\km$). The 
data for different $p_k$ were then combined, as follows: For each $p_k$, 
$P_k(N|\mu=0)$ was constructed from the transition matrix elements, then 
extrapolated to $p=\tilde{p}$ of interest using \eq{eq:vink}. The latter define 
the quantities $\Delta G_k(N) \equiv \ln \left(P_k(N)/P_k(N-1)\right)$, which 
were averaged over the $\km$ measurements
\begin{equation}\label{eq:av}
\begin{split}
 \Delta G(N) &= \frac{\sum_{k=1}^\km w_k \Delta G_k(N)} 
 {\sum_{k=1}^\km w_k} \quad, \\
 w_k &= \left( H_k(N-1)+H_k(N) \right) e^{ -|\tilde{p} - p_k| } \quad,
\end{split}
\end{equation}
where $H_k(N)$ counts how often the $k$-th simulation visited the state with $N$ 
particles (a simulation performed at $p_k$ is thus weighted by its 
\ahum{distance} from $\tilde{p}$, and the number of samples it contains). The 
distribution $P(N|\mu=0)$ is obtained via recursion:
\begin{equation}
 \ln P(N_{\rm min}) = 0, \quad
 \ln P(N) = \ln P(N-1) + \Delta G(N),
\end{equation}
$N_{\rm min}=\rmin L^2$, which can be extrapolated to different chemical 
potentials using \eq{eq:mu}. 

\section{Results}

\subsection{Locating the phase transition}

To scan the phase behavior of \eq{eq:vw}, we choose a value of the sharpness 
parameter $p$, and vary the chemical potential $\mu$. For small $p$, we expect a 
continuous KT transition, at some transition chemical potential $\mu_{\rm 
KT}$~\cite{fn1}. For large $p$, we expect a first-order transition, at chemical 
potential $\mu_{\rm 1st}$. For a tricritical point, the curves $\mu_{\rm KT}(p)$ 
and $\mu_{\rm 1st}(p)$ should form a single smooth line in the $(p,\mu)$-plane, 
i.e.~they should not cross or bifurcate.

The first-order transition is characterized by a density gap between the (then 
coexisting) isotropic and quasi-nematic phases~\cite{vink.wensink:2007}. To 
locate this transition, we introduce $\mustar$, defined as the chemical 
potential where the density fluctuation $\mean{N^2}-\mean{N}^2$ is maximized, as 
measured in a finite system of size $L$~\cite{orkoulas.fisher.ea:2001}. Here, 
$\mean{\cdot}$ is a grand canonical average, $\mean{N^a} = \sum_N N^a P(N) / 
\sum_N P(N)$, with $N_{\rm min} \leq N \leq N_{\rm max}$. In the thermodynamic 
limit, $L \to \infty$, the finite-size estimate $\mustar \to \mu_{\rm 1st}$, 
providing a means to locate the first-order transition.

The KT transition is characterized by diverging orientational 
fluctuations~\cite{citeulike:5931786}. Hence, we introduce $\mu^S_L$, defined as 
the chemical potential where the orientational fluctuation 
$\mean{S^2}-\mean{S}^2$ is maximized, again measured for finite $L$. Here, the 
nematic order parameter $S$ is the maximum eigenvalue of the 2D tensor 
$Q_{\alpha\beta} = (1/L^2) \sum_{i=1}^N 2 d_{i,\alpha} d_{i,\beta} - 
\delta_{\alpha,\beta}$~\cite{physreva.31.1776}, with the sum over all particles, 
$\delta$ the Kronecker-delta symbol, and $d_{i,\alpha}$ the $\alpha$-component 
of the vector $\vec{d}_i$ ($\alpha,\beta \in x,y)$. In the thermodynamic limit, 
$L \to \infty$, the finite-size estimate $\mu^S_L \to \mu_{\rm KT}$, providing a 
means to locate the KT transition.

\begin{figure}
\begin{center}
\includegraphics[width=0.8\columnwidth]{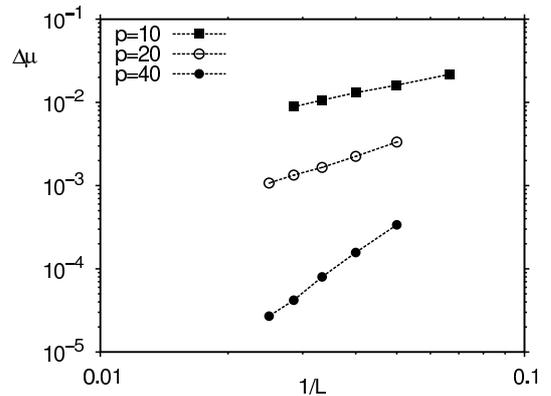}

\caption{\label{fig:locus} The chemical potential difference $\Delta \mu = 
\mustar-\mu^S_L$ versus $1/L$ on double-logarithmic scales, for several values 
of $p$ as indicated. For increasing $L$, $\Delta \mu$ decays as a power-law. The 
implication is that, in the thermodynamic limit, the line of KT transitions 
joins the line of first-order transitions.}

\end{center}
\end{figure}

\begin{figure}
\begin{center}
\includegraphics[width=0.8\columnwidth]{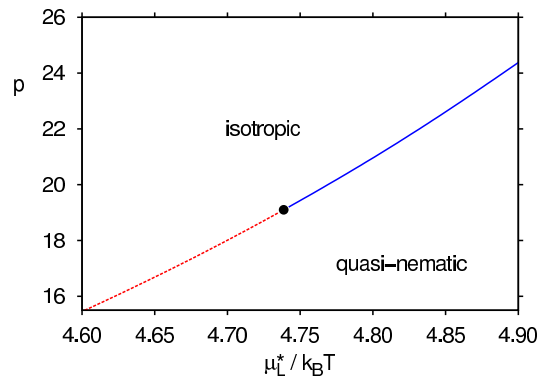}

\caption{\label{figpd} Phase diagram of \eq{eq:vw} in grand canonical 
representation, with the transition chemical potential $\mustar$ on the 
horizontal axes, and the exponent $p$ on the vertical one (data apply to $L=40$; 
on the scale of the graph, finite-size effects are small). The line separates 
isotropic from quasi-nematic phases. The isotropic $\leftrightarrow$ 
quasi-nematic transition is continuous and of the Kosterlitz-Thouless type when 
$p<\pc$ (dashed) and first-order (solid) when $p>\pc$. In the first-order 
regime, there is a density gap between the phases. The dot, at $\pc \approx 
19.1$, is the tricritical point obtained via finite-size scaling. At the 
tricritical point, also the radial distribution function $g(r)$ decays 
algebraically.}

\end{center}
\end{figure}

In \fig{fig:locus}, we plot $\Delta \mu = \mustar-\mu^S_L$ versus $1/L$, for 
several values of the sharpness parameter $p$. For the small value, $p=10$, the 
transition is of the KT type; for the large value, $p=40$, the transition is 
first-order; the value $p=20$ is close to the tricritical point, as we will show 
later. In finite systems $\mustar>\mu^S_L$, giving the impression of two 
separate transitions. However, $\Delta \mu$ decays to zero with increasing $L$. 
Hence, in the thermodynamic limit, the finite-size estimates $\mustar$ and 
$\mu^S_L$ are identical, i.e.~the statepoint where the density fluctuations are 
maximal coincides with the maximum of the orientational fluctuations. 

For each value of $p$, there is thus only one transition chemical potential, 
implying that the line of KT transitions joins the line of first-order 
transitions, as is required for a tricritical point. \fig{figpd} shows the phase 
diagram, i.e.~$\mustar$ versus $p$, which indeed yields a smooth curve. This 
curve separates the (low density) isotropic phase, from the (high density) 
quasi-nematic phase (it does not say anything about the nature of the transition 
between the phases; this is studied later). In what follows, we will base our 
analysis on the finite-size estimator $\mustar$.

\subsection{Structural properties of the bulk phases}

We now address the structural properties of the isotropic and quasi-nematic 
phase. As stated earlier, both phases are characterized by short-ranged 
positional order. To show this, we consider the static structure factor, 
$S(\vec{q}) = \mean{(1/N) | \sum_{i=1}^N \exp (\imath \vec{q} \cdot \vec{r}_i) 
\, |^2}$, with the sum over all particles, $\vec{r}_i$ the position of the 
$i$-th particle, wave vectors $\vec{q} = 2\pi(n_x,n_y)/L$ with integers $n_x$ 
and $n_y$, and $\mean{\cdot}$ an ensemble average (in what follows, we use the 
angular averaged $S(q)$, where $q=|\vec{q}|$). Note that $S(q)$ is the Fourier 
transform of the radial distribution function $g(r)$, so both these quantities 
contain the same information.

\begin{figure}
\begin{center}

\includegraphics[width=0.8\columnwidth]{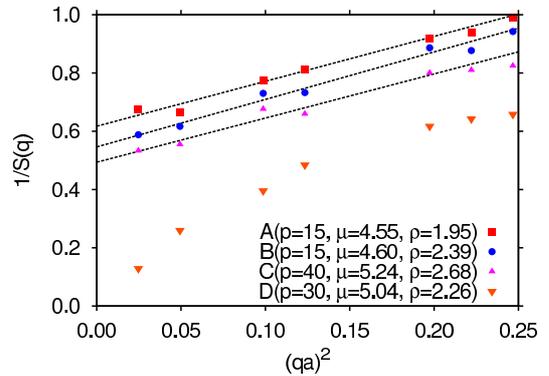}

\caption{\label{figsq} Static structure factor $1/S(q)$ versus $q^2$ in the 
limit $q \to 0$ for various statepoints $A-D$ as indicated (data apply to 
$L=40$). The statepoints $A-C$ are pure phases, taken well away from the 
transition line, corresponding to the isotropic phase $(A)$ and the 
quasi-nematic phase $(B,C)$. The dashed lines for $A-C$ are linear fits and 
confirm the Ornstein-Zernike form. The statepoint $D$ is taken on the phase 
transition line, using $p=\pg$ of the compressibility maximum. In this case, 
strong deviations from the Ornstein-Zernike formula are observed.}

\end{center}
\end{figure}

For chemical potentials $\mu$ away from the transition value $\mustar$, the $q 
\to 0$ limit of $S(q)$ is well described by the Ornstein-Zernike form, $1/S(q) = 
A(1 + \xi^2_r q^2)$, with $\xi_r$ the positional correlation length, and $A 
\equiv 1/S(0)=\mean{N} / (\mean{N^2}-\mean{N}^2)$~\cite{Hansen1986}. Some 
examples are shown in \fig{figsq} (statepoints $A-C$). The lines are linear 
fits, which for the correlation length yield typical values $\xi_r/a \sim 
1.6-1.7$, i.e.~short-ranged. Furthermore, the intercept of the fits is finite, 
$A>0$, which means that the density fluctuations are not diverging. Hence, as 
far as the positional order is concerned, the isotropic and quasi-nematic phase 
are both disordered fluids.

\begin{figure}
\begin{center}

\includegraphics[width=0.8\columnwidth]{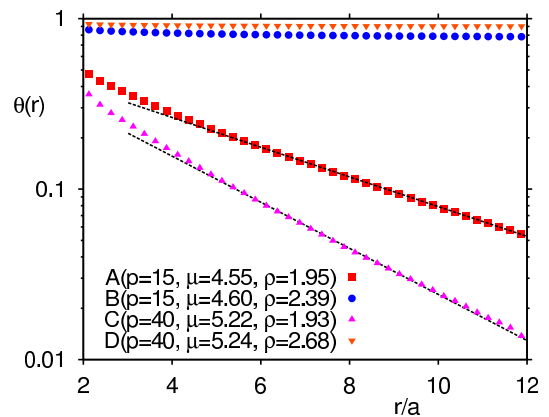}

\caption{\label{figtr} The orientational correlation function $\theta(r)$ for 
various statepoints $A-D$ as indicated (data apply to $L=40$; note the 
logarithmic vertical scale). All statepoints correspond to pure phases: 
isotropic $(A,C)$, and quasi-nematic $(B,D)$. In the isotropic phase, 
$\theta(r)$ decays exponentially; dashed lines show the corresponding fit. In 
the quasi-nematic phase, much slower (algebraic) decay is observed.}

\end{center}
\end{figure}

Next, we consider the orientational correlation function, $\theta(r) = \mean{2 
(\vec{d}_i \cdot \vec{d}_j)^2 - 1}'$~\cite{physreva.31.1776}, where 
$\mean{\cdot}'$ is an ensemble average over all pairs of particles $i-j$ for 
which $r_{ij}=r$ (in simulations, $\theta(r)$ is collected as a histogram). Some 
typical examples are shown in \fig{figtr}, where all the statepoints were chosen 
away from the phase transition. In the isotropic phase $(A,C)$, the 
orientational correlations decays exponentially, $\theta(r) \propto 
\exp(-r/\xi_\theta)$, with $\xi_\theta/a \sim 3-5$ obtained by fitting. In the 
quasi-nematic phase $(B,D)$, the decay is much slower, and best fitted with a 
power law, $\theta(r) \propto 1/r^{\eta_\theta}$, with $\eta_\theta$ being a 
small positive exponent. Hence, in the quasi-nematic phase, the orientational 
correlation length $\xi_\theta$ is infinite.

To summarize: The isotropic phase of \eq{eq:vw} is characterized by exponential 
decay of the positional and orientational correlations (both $\xi_r$ and 
$\xi_\theta$ being finite). In the quasi-nematic phase, the positional 
correlations still decay exponentially (finite $\xi_r$), while the orientational 
correlations decay algebraically $(\xi_\theta \to \infty)$.

\subsection{Nature of the phase transition}

\begin{figure}
\begin{center}
\includegraphics[width=\columnwidth]{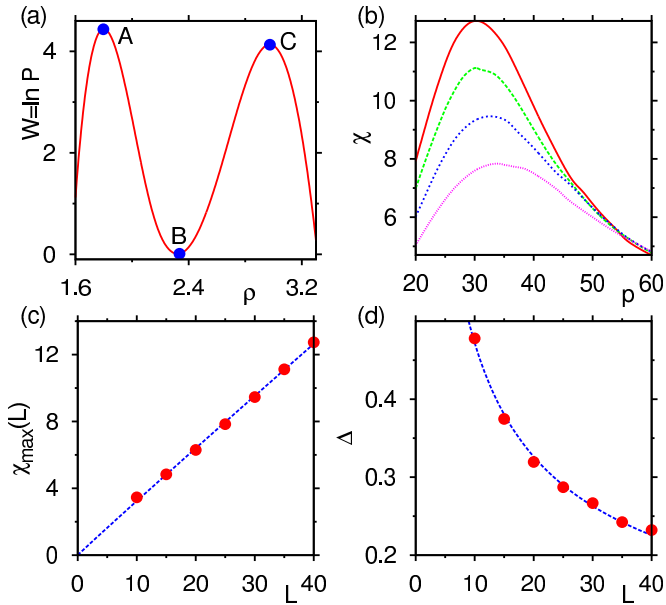}
\caption{\label{fig1} (a) Example distribution $W \equiv \ln P(\rho)$ for $p=75$ 
and $L=15$. The extrema $X(\rho_X,W_X)$, $X \in \{A,B,C\}$, define the order 
parameter $\Delta = \rho_C-\rho_A$, the coexistence diameter $\delta = 
(\rho_C+\rho_A)/2$, and the barrier $\Delta F = (W_A+W_C)/2-W_B$. (b) 
Susceptibility $\chi$ versus $p$ for $L=25,30,35,40$ (bottom to top). The curves 
reach finite maximum values $\xm$ at $p=\pg$. (c) Susceptibility 
maximum $\xm$ versus $L$. (d) Order parameter $\Delta$ at $p=\pg$ 
versus $L$. The dashed curves in (c,d) are power law fits.}
\end{center}
\end{figure}

We now consider the nature of the isotropic $\leftrightarrow$ quasi-nematic 
transition, and how the transition type changes with the sharpness parameter 
$p$. To this end, we follow the path $p(\mustar)$ in the phase diagram of 
\fig{figpd}, and record how the distribution $P(N)$, and the quantities derived 
from it, vary along it (i.e.~for each value of $p$, the chemical potential is 
tuned such that the variance in the particle number is maximized). For large 
$p$, where the transition is strongly first-order~\cite{vink.wensink:2007}, 
$P(N)$ is bimodal. An example is shown in \fig{fig1}(a). The presence of two 
peaks implies two-phase coexistence (to this end, it may be useful to interpret 
{\it minus} $\ln P(N)$ as the free energy of the system). The left (right) peak 
corresponds to the isotropic (quasi-nematic) phase. The distance between the 
peaks reflects the density gap between the phases, which we take as the order 
parameter $\Delta$ of the transition. It is numerically convenient to compute 
the order parameter as $\Delta = \mean{|M|}/L^2$, $M=N-\mean{N}$. Similarly, we 
introduce the order parameter fluctuations (susceptibility) $\chi = \left( 
\mean{M^2} - \mean{|M|}^2 \right)/L^2$~\cite{orkoulas.panagiotopoulos.ea:2000}.

At the tricritical point, $p=\pc$, the density gap $\Delta$ vanishes. To locate 
this point, we perform a finite-size scaling analysis. \fig{fig1}(b) shows $\chi$ 
versus $p$ for several $L$. We note that each curve reveals a maximum. The value 
of $p$ at the maximum defines $\pg$, the corresponding value of the 
susceptibility defines $\xm$ (we emphasize that both these quantities are 
$L$-dependent). The fact that $\xm$ increases with $L$ indicates that, at the 
tricritical point $\pc = \lim_{L \to \infty} \pg$, the susceptibility diverges. 
We observe a power-law increase, $\xm \propto L^{\omega_1}$, with $\omega_1 
\approx 1.0 \pm 0.03$ obtained by fitting [\fig{fig1}(c)]. For the order 
parameter, measured at $p=\pg$, we observe a power-law decay, $\Delta \propto 
L^{-\omega_2}$, where a fit yields $\omega_2 \approx 0.5 \pm 0.03$ 
[\fig{fig1}(d)]. Note that the exponents obey hyperscaling, 
$\omega_1+2\omega_2=d=2$, as is characteristic of critical and tricritical 
transitions~\cite{citeulike:9070838}. This implies that, at the tricritical 
point, the distribution $P(N)$ is scale invariant.

\begin{figure}
\begin{center}
\includegraphics[width=\columnwidth]{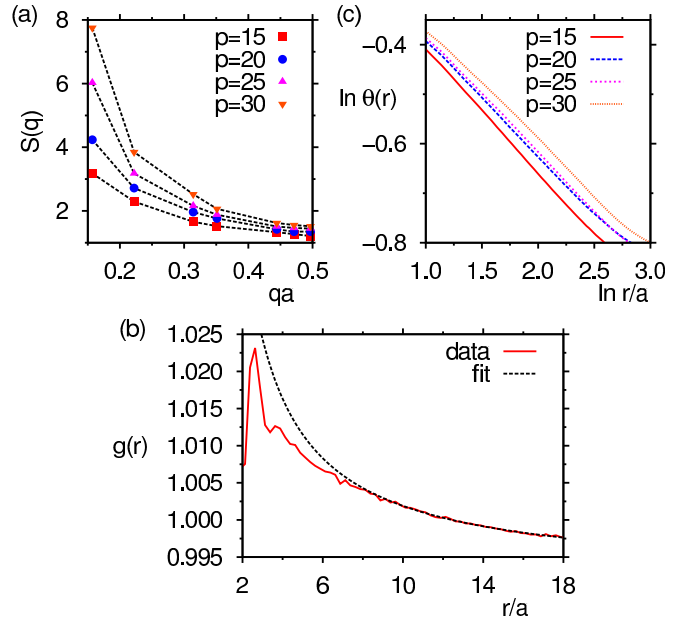}

\caption{\label{fig:sqpath} Structural properties measured along the path 
$p(\mustar)$, i.e.~the line of isotropic $\leftrightarrow$ quasi-nematic phase 
transitions of \fig{figpd}, restricted to the range $p \leq \pg$. The data apply 
to $L=40$, for which $\pg \sim 30$. (a) The static structure factor $S(q)$ for 
various values of $p$ as indicated. As $p \to \pg$, $S(q \to 0)$ strongly 
increases, consistent with a critical point at $\pg$. (b) The radial 
distribution function $g(r)$ at $\pg$, together with a fit to \eq{eq:gr}, the 
fit range being $10<r/a<18$. (c) The orientational correlation function 
$\theta(r)$ for several values of $p$. Note the double-logarithmic scale! The 
decay of $\theta(r)$ is algebraic.}

\end{center}
\end{figure}

The diverging susceptibility is also manifested by the static structure factor 
measured along the path $p(\mustar)$. As $p \to \pg$, $S(q \to 0)$ strongly 
increases, consistent with diverging order parameter fluctuations 
[\fig{fig:sqpath}(a)]. Note that in \fig{fig:sqpath} the tricritical point is 
approached from below, i.e.~starting with small $p$. This was done for 
convenience: Approaching the tricritical point from above would require $S(q)$ 
to be measured for the isotropic and quasi-nematic phase separately, since these 
phases coexist when $p>\pg$. At the tricritical point, $S(q)$ strongly deviates 
from the Ornstein-Zernike formula, with $1/S(q \to 0)$ now tending to zero 
[\fig{figsq}, statepoint~$D$]. A diverging susceptibility implies that, at the 
tricritical point, also the positional correlations decay algebraically, 
i.e.~$\xi_r \to \infty$. In 2D, the radial distribution function should then 
decay asymptotically as~\cite{citeulike:8828115}
\begin{equation} \label{eq:gr}
 \lim_{r \to \infty} g(r)=c_1 + c_2/r^{\eta_r} \quad, 
\end{equation}
with $\eta_r=2-\omega_1 \sim 1.0$, and constants $c_i$. \fig{fig:sqpath}(b) 
shows that $g(r)$ at $\pg$ is indeed well described by this form, where 
$\eta_r=1$ was imposed, and the constants $c_i$ were fitted.

In \fig{fig:sqpath}(c), we plot the orientational correlation function 
$\theta(r)$ measured along the path $p(\mustar)$. All along the path 
$p(\mustar)$, $\theta(r)$ decays algebraically. At $\pg$, the exponent of the 
algebraic decay of the orientational correlations $\eta_\theta \sim 0.22$, 
i.e.~much slower than the decay of the positional correlations. In contrast, the 
radial distribution function $g(r)$ decays algebraically only at the 
tricritical point. The simultaneous divergence of two order parameter 
fluctuations (here: density and orientation), implied by the algebraic decay of 
the corresponding correlation functions, is characteristic of tricritical 
phenomena.

\begin{figure}
\begin{center}
\includegraphics[width=\columnwidth]{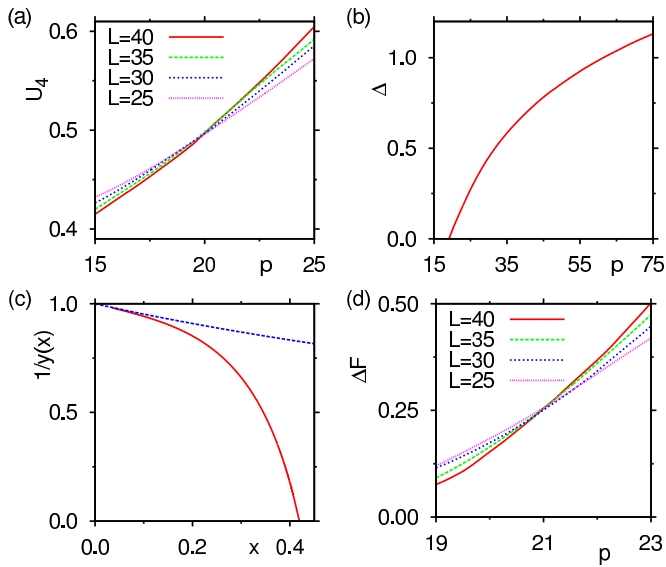}
\caption{\label{fig2} (a) Binder cumulant $U_4$ versus $p$ for several $L$. The 
intersections mark $\pc$. (b) Order parameter $\Delta$ versus $p$ obtained 
following Kim and Fisher~\cite{kim.fisher:2004}. At $\pc$, the order parameter 
vanishes. (c) The (inverse) scaling function $y(x)$ of the Kim-Fisher algorithm 
(solid). Also shown is the $x \to 0$ limiting form (dashed), which our data 
indeed approach. (d) The barrier $\Delta F$ versus $p$ for several $L$. At the 
tricritical point, the curves for different $L$ intersect.}
\end{center}
\end{figure}

\subsection{Determination of $\pc$}

Finally, we determine $\pc$. The standard approach is to consider the Binder 
cumulant $U_4 = \mean{M^2}^2 / \mean{M^4}$; owing to hyperscaling, the latter is 
$L$-independent at $\pc$~\cite{binder:1981}. In \fig{fig2}(a), we plot $U_4$ 
versus $p$ for various $L$. We observe a scatter of intersections, between 
$18.9<p<20.4$, providing a rough estimate of $\pc$ (corrections to scaling 
appear to be quite strong, and so we restrict the analysis to the largest four 
system sizes in what follows). A more precise estimate of $\pc$ is obtained 
using the complete scaling algorithm of Kim and Fisher~\cite{kim.fisher:2004}. 
For the practical implementation of the latter, our $p$-extrapolation scheme, 
i.e.~\eq{eq:vink}, is absolutely crucial, since data over a wide range in $p$ 
are required (stretching from the first-order to the tricritical regime). The 
principal output of the complete scaling algorithm is the $L \to \infty$ value 
of the order parameter $\Delta$ as a function of $p$ [\fig{fig2}(b)]. From this, 
we conclude $\pc = 19.1 \pm 0.1$, i.e.~the value where $\Delta$ vanishes. Note 
that this estimate is consistent with the cumulant intersections.

A second output of the complete scaling algorithm is a scaling function $y(x)$, 
defined in the Appendix, which is characteristic of the universality class 
[\fig{fig2}(c)]. In the limit $x \to 0$, $y(x)=1+x/2$, while at some $x_{\rm 
c}>0$, $y(x)$ diverges. We obtain $x_{\rm c} \approx 0.42$. As a last method to 
obtain $\pc$, we consider the barrier $\Delta F$ of $\ln P(N)$, defined in 
\fig{fig1}(a) as the average height of the peaks (A and B) minus the height at 
the minimum (C). The barrier increases (decreases) with $L$ for $p>\pc$ 
($p<\pc$), and remains $L$-independent at $\pc$~\cite{binder:1982, 
physrevlett.65.137}. The variation of $\Delta F$ with $p$ for several $L$ is 
shown in \fig{fig2}(d). At the tricritical point, the curves for different $L$ 
intersect, at values of $p$ consistent with those of the cumulant analysis.

\begin{figure}
\begin{center}
\includegraphics[width=\columnwidth]{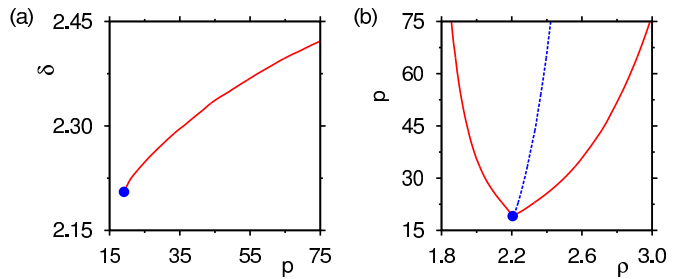}
\caption{\label{fig3} (a) Coexistence diameter $\delta$ versus $p$ obtained 
following \olcite{kim:2005}. (b) $L \to \infty$ phase diagram of \eq{eq:vw}, 
showing the binodal (solid) and diameter (dashed). The tricritical 
point (dots) is at $\pc=19.1$ and $\rho_{\rm c}=2.205$.}
\end{center}
\end{figure}

\subsection{Phase diagram in $(\rho,p)$-representation}

For completeness, we still compute the $L \to \infty$ phase diagram in 
$(p,\rho)$-representation. Kim also provides a scaling algorithm to obtain the 
$L \to \infty$ coexistence diameter $\delta$ from finite-size simulation 
data~\cite{kim:2005}. The latter is defined as the average density of the 
isotropic and quasi-nematic phase [\fig{fig1}(a)]. In \fig{fig3}(a), we plot 
$\delta$ versus $p$. The order parameter and coexistence diameter yield the 
binodal, i.e.~the density of the isotropic ($\delta-\Delta/2$) and quasi-nematic 
phase ($\delta+\Delta/2$) as a function of $p$ [\fig{fig3}(b)]. The region 
inside the binodal marks the statepoints where both these phases coexist. Note 
that the isotropic and quasi-nematic branches form a \ahum{kink} at the 
tricritical point, in agreement with a mean-field treatment of 
\eq{eq:vw}~\cite{vink.wensink:2007}. Not shown in the phase diagrams of 
\fig{fig3} is the line of continuous KT transitions that commence below the 
tricritical point.

\section{Discussion and summary}

In summary, we have considered the crossover of the Kosterliz-Thouless 
transition in 2D liquid crystals from continuous to first-order. Our main result 
is that, at the crossover, a tricritical point occurs. At the tricritical 
point, both the positional and orientational correlations decay algebraically. 
The algebraic decay of positional order enhances the spectrum of possible 
structure in 2D liquid crystals, since positional order in quasi-nematic phases 
is typically assumed to decay exponentially. 

It may be that the tricritical point we found is universal, in the sense that 
any model with sufficiently sharp interactions and 2D positional/vector degrees 
of freedom would yield the same set of tricritical exponents, $\omega_1$ and 
$\omega_2$. To test this hypothesis, it would be interesting to apply the 
analysis of this work to lattice-based models, such as the one studied by Domany 
and co-workers~\cite{physrevlett.52.1535}. In that case, the analysis could be 
based on the energy distribution $P(E)$, which also becomes bimodal when the 
transition is first-order. Such an analysis is furthermore interesting because 
there is not yet consensus about how the first-order transition ends. The 
simultaneous divergence of the density and orientational fluctuations observed 
by us indicates a tricritical point, while studies of lattice-based models also 
report critical point behavior~\cite{physrevlett.70.1327}. According to 
\olcite{citeulike:12935130}, in 2D spatial dimensions, lowering $p$ leads to a 
2D Ising critical point, but this assumes the absence of a KT 
transition~\cite{fn2}. In agreement with this, using 2D spatial dimensions and 
3D vector spins (Heisenberg case), a KT transition is not expected 
(orientational correlations always decay exponentially). In that case, numerical 
simulations~\cite{physrevlett.88.047203} are consistent with a 2D Ising critical 
point, i.e.~$\omega_1=7/4$ and $\omega_2=1/8$.

The present analysis was largely facilitated by a method to extrapolate 
simulation data in the sharpness parameter~$p$. However, it is by no means 
restricted to the model of \eq{eq:vw}, and can be applied to any variable in any 
potential, provided an explicit expression for the expansion \eq{eq:vink} can be 
given. In particular, it can also be used to extrapolate in field variables, 
i.e.~the type of variables (temperature, chemical potential) for which histogram 
reweighting~\cite{ferrenberg.swendsen:1988} was originally intended. Due to its 
modest storage requirements, our scheme could prove attractive even then. For an 
explicit demonstration, we refer the reader to the Appendix of 
\olcite{Vink2014}, where extrapolations in temperature are performed in this 
manner. For the future, it would be useful to develop a more rigorous version of 
\eq{eq:av} to combine data obtained for different values of the control 
parameters, along the lines of the multiple-histogram 
method~\cite{ferrenberg.swendsen:1989}.

\acknowledgments

Financial support from the Emmy Noether program (grant number: VI~483) of the 
German Research Foundation is acknowledged. I also thank anonymous referees for 
pointing out the possibility of tricritical behavior, as well as the need to 
study in detail the orientational correlations. In addition, I thank A.~van 
Enter for useful discussions.

\bibliography{refs_VINK,notes}

\begin{thebibliography}{40}
\expandafter\ifx\csname natexlab\endcsname\relax\def\natexlab#1{#1}\fi
\expandafter\ifx\csname bibnamefont\endcsname\relax
  \def\bibnamefont#1{#1}\fi
\expandafter\ifx\csname bibfnamefont\endcsname\relax
  \def\bibfnamefont#1{#1}\fi
\expandafter\ifx\csname citenamefont\endcsname\relax
  \def\citenamefont#1{#1}\fi
\expandafter\ifx\csname url\endcsname\relax
  \def\url#1{\texttt{#1}}\fi
\expandafter\ifx\csname urlprefix\endcsname\relax\def\urlprefix{URL }\fi
\providecommand{\bibinfo}[2]{#2}
\providecommand{\eprint}[2][]{\url{#2}}

\bibitem[{\citenamefont{Wittebrood et~al.}(1996)\citenamefont{Wittebrood,
  Luijendijk, Stallinga, Rasing, and Mu\v{s}evi\v{c}}}]{citeulike:4006159}
\bibinfo{author}{\bibfnamefont{M.}~\bibnamefont{Wittebrood}},
  \bibinfo{author}{\bibfnamefont{D.}~\bibnamefont{Luijendijk}},
  \bibinfo{author}{\bibfnamefont{S.}~\bibnamefont{Stallinga}},
  \bibinfo{author}{\bibfnamefont{T.}~\bibnamefont{Rasing}}, \bibnamefont{and}
  \bibinfo{author}{\bibfnamefont{I.}~\bibnamefont{Mu\v{s}evi\v{c}}},
  \bibinfo{journal}{Phys.~Rev. E} \textbf{\bibinfo{volume}{54}},
  \bibinfo{pages}{5232} (\bibinfo{year}{1996}), ISSN \bibinfo{issn}{1063-651X},
  \urlprefix\url{http://dx.doi.org/10.1103/physreve.54.5232}.

\bibitem[{\citenamefont{Garcia et~al.}(2008)\citenamefont{Garcia, Subashi, and
  Fukuto}}]{citeulike:2811025}
\bibinfo{author}{\bibfnamefont{R.}~\bibnamefont{Garcia}},
  \bibinfo{author}{\bibfnamefont{E.}~\bibnamefont{Subashi}}, \bibnamefont{and}
  \bibinfo{author}{\bibfnamefont{M.}~\bibnamefont{Fukuto}},
  \bibinfo{journal}{Phys.~Rev. Lett.} \textbf{\bibinfo{volume}{100}},
  \bibinfo{pages}{197801} (\bibinfo{year}{2008}), ISSN
  \bibinfo{issn}{0031-9007},
  \urlprefix\url{http://dx.doi.org/10.1103/physrevlett.100.197801}.

\bibitem[{\citenamefont{van Effenterre et~al.}(2001)\citenamefont{van
  Effenterre, Ober, Valignat, and Cazabat}}]{citeulike:3991276}
\bibinfo{author}{\bibfnamefont{D.}~\bibnamefont{van Effenterre}},
  \bibinfo{author}{\bibfnamefont{R.}~\bibnamefont{Ober}},
  \bibinfo{author}{\bibfnamefont{M.~P.} \bibnamefont{Valignat}},
  \bibnamefont{and} \bibinfo{author}{\bibfnamefont{A.~M.}
  \bibnamefont{Cazabat}}, \bibinfo{journal}{Phys.~Rev. Lett.}
  \textbf{\bibinfo{volume}{87}}, \bibinfo{pages}{125701}
  (\bibinfo{year}{2001}), ISSN \bibinfo{issn}{0031-9007},
  \urlprefix\url{http://dx.doi.org/10.1103/physrevlett.87.125701}.

\bibitem[{\citenamefont{Jordens et~al.}(2013)\citenamefont{Jordens, Isa, Usov,
  and Mezzenga}}]{Jordens2013}
\bibinfo{author}{\bibfnamefont{S.}~\bibnamefont{Jordens}},
  \bibinfo{author}{\bibfnamefont{L.}~\bibnamefont{Isa}},
  \bibinfo{author}{\bibfnamefont{I.}~\bibnamefont{Usov}}, \bibnamefont{and}
  \bibinfo{author}{\bibfnamefont{R.}~\bibnamefont{Mezzenga}},
  \bibinfo{journal}{Nature Communications} \textbf{\bibinfo{volume}{4}},
  \bibinfo{pages}{1917} (\bibinfo{year}{2013}), ISSN \bibinfo{issn}{2041-1723},
  \urlprefix\url{http://dx.doi.org/10.1038/ncomms2911}.

\bibitem[{\citenamefont{Straley}(1971)}]{physreva.4.675}
\bibinfo{author}{\bibfnamefont{J.~P.} \bibnamefont{Straley}},
  \bibinfo{journal}{Phys. Rev. A} \textbf{\bibinfo{volume}{4}},
  \bibinfo{pages}{675} (\bibinfo{year}{1971}),
  \urlprefix\url{http://dx.doi.org/10.1103/PhysRevA.4.675}.

\bibitem[{\citenamefont{Frenkel and Eppenga}(1985)}]{physreva.31.1776}
\bibinfo{author}{\bibfnamefont{D.}~\bibnamefont{Frenkel}} \bibnamefont{and}
  \bibinfo{author}{\bibfnamefont{R.}~\bibnamefont{Eppenga}},
  \bibinfo{journal}{Phys. Rev. A} \textbf{\bibinfo{volume}{31}},
  \bibinfo{pages}{1776} (\bibinfo{year}{1985}),
  \urlprefix\url{http://dx.doi.org/10.1103/PhysRevA.31.1776}.

\bibitem[{\citenamefont{Bates and Frenkel}(2000)}]{bates.frenkel:2000}
\bibinfo{author}{\bibfnamefont{M.~A.} \bibnamefont{Bates}} \bibnamefont{and}
  \bibinfo{author}{\bibfnamefont{D.}~\bibnamefont{Frenkel}},
  \bibinfo{journal}{J.~Chem.~Phys.} \textbf{\bibinfo{volume}{112}},
  \bibinfo{pages}{10034} (\bibinfo{year}{2000}),
  \urlprefix\url{http://dx.doi.org/10.1063/1.481637}.

\bibitem[{\citenamefont{Vink}(2009)}]{citeulike:5931786}
\bibinfo{author}{\bibfnamefont{R.~L.~C.} \bibnamefont{Vink}},
  \bibinfo{journal}{Eur. Phys. J. B} \textbf{\bibinfo{volume}{72}},
  \bibinfo{pages}{225} (\bibinfo{year}{2009}), ISSN \bibinfo{issn}{1434-6036},
  \urlprefix\url{http://dx.doi.org/10.1140/epjb/e2009-00333-x}.

\bibitem[{\citenamefont{Lagomarsino et~al.}(2003)\citenamefont{Lagomarsino,
  Dogterom, and Dijkstra}}]{lagomarsino.dogterom.ea:2003}
\bibinfo{author}{\bibfnamefont{M.~C.} \bibnamefont{Lagomarsino}},
  \bibinfo{author}{\bibfnamefont{M.}~\bibnamefont{Dogterom}}, \bibnamefont{and}
  \bibinfo{author}{\bibfnamefont{M.}~\bibnamefont{Dijkstra}},
  \bibinfo{journal}{J.~Chem.~Phys.} \textbf{\bibinfo{volume}{119}},
  \bibinfo{pages}{3535} (\bibinfo{year}{2003}),
  \urlprefix\url{http://dx.doi.org/10.1063/1.1588994}.

\bibitem[{\citenamefont{Kosterlitz and
  Thouless}(1973)}]{kosterlitz.thouless:1973}
\bibinfo{author}{\bibfnamefont{J.~M.} \bibnamefont{Kosterlitz}}
  \bibnamefont{and} \bibinfo{author}{\bibfnamefont{D.~J.}
  \bibnamefont{Thouless}}, \bibinfo{journal}{J.~Phys.~C}
  \textbf{\bibinfo{volume}{6}}, \bibinfo{pages}{1181} (\bibinfo{year}{1973}),
  ISSN \bibinfo{issn}{0022-3719},
  \urlprefix\url{http://dx.doi.org/10.1088/0022-3719/6/7/010}.

\bibitem[{\citenamefont{Kosterlitz}(1974)}]{kosterlitz:1974}
\bibinfo{author}{\bibfnamefont{J.~M.} \bibnamefont{Kosterlitz}},
  \bibinfo{journal}{J.~Phys.~C} \textbf{\bibinfo{volume}{7}},
  \bibinfo{pages}{1046} (\bibinfo{year}{1974}), ISSN \bibinfo{issn}{0022-3719},
  \urlprefix\url{http://dx.doi.org/10.1088/0022-3719/7/6/005}.

\bibitem[{\citenamefont{Yokoyama}(1988)}]{citeulike:4067023}
\bibinfo{author}{\bibfnamefont{H.}~\bibnamefont{Yokoyama}},
  \bibinfo{journal}{J. Chem. Soc., Faraday Trans. 2}
  \textbf{\bibinfo{volume}{84}}, \bibinfo{pages}{1023} (\bibinfo{year}{1988}),
  \urlprefix\url{http://dx.doi.org/10.1039/f29888401023}.

\bibitem[{\citenamefont{van Enter and Shlosman}(2002)}]{physrevlett.89.285702}
\bibinfo{author}{\bibfnamefont{A.~C.~D.} \bibnamefont{van Enter}}
  \bibnamefont{and} \bibinfo{author}{\bibfnamefont{S.~B.}
  \bibnamefont{Shlosman}}, \bibinfo{journal}{Phys.~Rev. Lett.}
  \textbf{\bibinfo{volume}{89}}, \bibinfo{pages}{285702}
  (\bibinfo{year}{2002}),
  \urlprefix\url{http://dx.doi.org/10.1103/physrevlett.89.285702}.

\bibitem[{\citenamefont{Enter and Shlosman}(2005)}]{enter.shlosman:2005}
\bibinfo{author}{\bibfnamefont{A.~C.} \bibnamefont{Enter}} \bibnamefont{and}
  \bibinfo{author}{\bibfnamefont{S.~B.} \bibnamefont{Shlosman}},
  \bibinfo{journal}{Communications in Mathematical Physics}
  \textbf{\bibinfo{volume}{255}}, \bibinfo{pages}{21} (\bibinfo{year}{2005}),
  ISSN \bibinfo{issn}{0010-3616},
  \urlprefix\url{http://dx.doi.org/10.1007/s00220-004-1286-1}.

\bibitem[{\citenamefont{van Enter and Shlosman}(2007)}]{citeulike:12935130}
\bibinfo{author}{\bibfnamefont{A.}~\bibnamefont{van Enter}} \bibnamefont{and}
  \bibinfo{author}{\bibfnamefont{S.}~\bibnamefont{Shlosman}},
  \bibinfo{journal}{Markov Processes Relat. Fields}
  \textbf{\bibinfo{volume}{13}}, \bibinfo{pages}{239} (\bibinfo{year}{2007}),
  \urlprefix\url{http://arxiv.org/abs/cond-mat/0506730}.

\bibitem[{\citenamefont{Wensink and Vink}(2007)}]{vink.wensink:2007}
\bibinfo{author}{\bibfnamefont{H.~H.} \bibnamefont{Wensink}} \bibnamefont{and}
  \bibinfo{author}{\bibfnamefont{R.~L.~C.} \bibnamefont{Vink}},
  \bibinfo{journal}{J.~Phys.:~Condens.~Matter} \textbf{\bibinfo{volume}{19}},
  \bibinfo{pages}{466109} (\bibinfo{year}{2007}),
  \urlprefix\url{http://stacks.iop.org/0953-8984/19/466109}.

\bibitem[{\citenamefont{Lawrie and Sarbach}(1984)}]{Lawrie1984}
\bibinfo{author}{\bibfnamefont{I.~D.} \bibnamefont{Lawrie}} \bibnamefont{and}
  \bibinfo{author}{\bibfnamefont{S.}~\bibnamefont{Sarbach}}, in
  \emph{\bibinfo{booktitle}{Phase Transitions and Critical Phenomena}}, edited
  by \bibinfo{editor}{\bibfnamefont{C.}~\bibnamefont{Domb}} \bibnamefont{and}
  \bibinfo{editor}{\bibfnamefont{J.~L.} \bibnamefont{Lebowitz}}
  (\bibinfo{publisher}{Academic Press}, \bibinfo{address}{London},
  \bibinfo{year}{1984}), vol.~\bibinfo{volume}{9}, chap.~\bibinfo{chapter}{1},
  p.~\bibinfo{pages}{1}.

\bibitem[{\citenamefont{Ferrenberg and
  Swendsen}(1988)}]{ferrenberg.swendsen:1988}
\bibinfo{author}{\bibfnamefont{A.~M.} \bibnamefont{Ferrenberg}}
  \bibnamefont{and} \bibinfo{author}{\bibfnamefont{R.~H.}
  \bibnamefont{Swendsen}}, \bibinfo{journal}{Phys.~Rev. Lett.}
  \textbf{\bibinfo{volume}{61}}, \bibinfo{pages}{2635} (\bibinfo{year}{1988}),
  \urlprefix\url{http://dx.doi.org/10.1103/physrevlett.61.2635}.

\bibitem[{\citenamefont{Frenkel and Smit}(2001)}]{frenkel.smit:2001}
\bibinfo{author}{\bibfnamefont{D.}~\bibnamefont{Frenkel}} \bibnamefont{and}
  \bibinfo{author}{\bibfnamefont{B.}~\bibnamefont{Smit}},
  \emph{\bibinfo{title}{Understanding Molecular Simulation}}
  (\bibinfo{publisher}{Academic Press}, \bibinfo{address}{San Diego},
  \bibinfo{year}{2001}).

\bibitem[{\citenamefont{Shell et~al.}(2003)\citenamefont{Shell, Debenedetti,
  and Panagiotopoulos}}]{citeulike:3577799}
\bibinfo{author}{\bibfnamefont{M.~S.} \bibnamefont{Shell}},
  \bibinfo{author}{\bibfnamefont{P.~G.} \bibnamefont{Debenedetti}},
  \bibnamefont{and} \bibinfo{author}{\bibfnamefont{A.~Z.}
  \bibnamefont{Panagiotopoulos}}, \bibinfo{journal}{J.~Chem.~Phys.}
  \textbf{\bibinfo{volume}{119}}, \bibinfo{pages}{9406} (\bibinfo{year}{2003}),
  \urlprefix\url{http://dx.doi.org/10.1063/1.1615966}.

\bibitem[{\citenamefont{Wang and Landau}(2001)}]{wang.landau:2001}
\bibinfo{author}{\bibfnamefont{F.}~\bibnamefont{Wang}} \bibnamefont{and}
  \bibinfo{author}{\bibfnamefont{D.~P.} \bibnamefont{Landau}},
  \bibinfo{journal}{Phys.~Rev. Lett.} \textbf{\bibinfo{volume}{86}},
  \bibinfo{pages}{2050} (\bibinfo{year}{2001}),
  \urlprefix\url{http://dx.doi.org/10.1103/physrevlett.86.2050}.

\bibitem[{\citenamefont{Fitzgerald et~al.}(1999)\citenamefont{Fitzgerald,
  Picard, and Silver}}]{citeulike:12476499}
\bibinfo{author}{\bibfnamefont{M.}~\bibnamefont{Fitzgerald}},
  \bibinfo{author}{\bibfnamefont{R.~R.} \bibnamefont{Picard}},
  \bibnamefont{and} \bibinfo{author}{\bibfnamefont{R.~N.}
  \bibnamefont{Silver}}, \bibinfo{journal}{EPL} p. \bibinfo{pages}{282}
  (\bibinfo{year}{1999}),
  \urlprefix\url{http://dx.doi.org/10.1209/epl/i1999-00257-1}.

\bibitem[{\citenamefont{Errington}(2003)}]{citeulike:3948984}
\bibinfo{author}{\bibfnamefont{J.~R.} \bibnamefont{Errington}},
  \bibinfo{journal}{Phys.~Rev. E} \textbf{\bibinfo{volume}{67}},
  \bibinfo{pages}{012102} (\bibinfo{year}{2003}),
  \urlprefix\url{http://dx.doi.org/10.1103/physreve.67.012102}.

\bibitem[{fn1()}]{fn1}
\bibinfo{note}{For very small $p$, the KT transition will eventually vanish,
  which might give rise to other interesting effects. This part of the phase
  diagram is not considered in this work.}

\bibitem[{\citenamefont{Orkoulas et~al.}(2001)\citenamefont{Orkoulas, Fisher,
  and Panagiotopoulos}}]{orkoulas.fisher.ea:2001}
\bibinfo{author}{\bibfnamefont{G.}~\bibnamefont{Orkoulas}},
  \bibinfo{author}{\bibfnamefont{M.~E.} \bibnamefont{Fisher}},
  \bibnamefont{and} \bibinfo{author}{\bibfnamefont{A.~Z.}
  \bibnamefont{Panagiotopoulos}}, \bibinfo{journal}{Phys.~Rev. E}
  \textbf{\bibinfo{volume}{63}}, \bibinfo{pages}{051507}
  (\bibinfo{year}{2001}),
  \urlprefix\url{http://dx.doi.org/10.1103/physreve.63.051507}.

\bibitem[{\citenamefont{Hansen and McDonald}(1986)}]{Hansen1986}
\bibinfo{author}{\bibfnamefont{J.~P.} \bibnamefont{Hansen}} \bibnamefont{and}
  \bibinfo{author}{\bibfnamefont{I.~R.} \bibnamefont{McDonald}},
  \emph{\bibinfo{title}{Theory of simple liquids}}
  (\bibinfo{publisher}{Academic}, \bibinfo{address}{London, New York},
  \bibinfo{year}{1986}), \bibinfo{edition}{2nd} ed.

\bibitem[{\citenamefont{Orkoulas et~al.}(2000)\citenamefont{Orkoulas,
  Panagiotopoulos, and Fisher}}]{orkoulas.panagiotopoulos.ea:2000}
\bibinfo{author}{\bibfnamefont{G.}~\bibnamefont{Orkoulas}},
  \bibinfo{author}{\bibfnamefont{A.~Z.} \bibnamefont{Panagiotopoulos}},
  \bibnamefont{and} \bibinfo{author}{\bibfnamefont{M.~E.}
  \bibnamefont{Fisher}}, \bibinfo{journal}{Phys. Rev. E}
  \textbf{\bibinfo{volume}{61}}, \bibinfo{pages}{5930} (\bibinfo{year}{2000}),
  \urlprefix\url{http://dx.doi.org/10.1103/PhysRevE.61.5930}.

\bibitem[{\citenamefont{Wilding and Nielaba}(1996)}]{citeulike:9070838}
\bibinfo{author}{\bibfnamefont{N.~B.} \bibnamefont{Wilding}} \bibnamefont{and}
  \bibinfo{author}{\bibfnamefont{P.}~\bibnamefont{Nielaba}},
  \bibinfo{journal}{Phys.~Rev. E} \textbf{\bibinfo{volume}{53}},
  \bibinfo{pages}{926} (\bibinfo{year}{1996}), ISSN \bibinfo{issn}{1063-651X},
  \urlprefix\url{http://dx.doi.org/10.1103/physreve.53.926}.

\bibitem[{\citenamefont{Pelissetto and Vicari}(2002)}]{citeulike:8828115}
\bibinfo{author}{\bibfnamefont{A.}~\bibnamefont{Pelissetto}} \bibnamefont{and}
  \bibinfo{author}{\bibfnamefont{E.}~\bibnamefont{Vicari}},
  \bibinfo{journal}{Phys.~Rep.} \textbf{\bibinfo{volume}{368}},
  \bibinfo{pages}{549} (\bibinfo{year}{2002}), ISSN \bibinfo{issn}{03701573},
  \eprint{cond-mat/0012164},
  \urlprefix\url{http://dx.doi.org/10.1016/s0370-1573(02)00219-3}.

\bibitem[{\citenamefont{Kim and Fisher}(2005)}]{kim.fisher:2004}
\bibinfo{author}{\bibfnamefont{Y.~C.} \bibnamefont{Kim}} \bibnamefont{and}
  \bibinfo{author}{\bibfnamefont{M.~E.} \bibnamefont{Fisher}},
  \bibinfo{journal}{Comput. Phys. Commun.} \textbf{\bibinfo{volume}{169}},
  \bibinfo{pages}{295} (\bibinfo{year}{2005}),
  \urlprefix\url{http://xxx.lanl.gov/abs/cond-mat/0411736}.

\bibitem[{\citenamefont{Binder}(1981)}]{binder:1981}
\bibinfo{author}{\bibfnamefont{K.}~\bibnamefont{Binder}}, \bibinfo{journal}{Z.
  Phys. B} \textbf{\bibinfo{volume}{43}}, \bibinfo{pages}{119}
  (\bibinfo{year}{1981}), ISSN \bibinfo{issn}{0340-224X},
  \urlprefix\url{http://dx.doi.org/10.1007/bf01293604}.

\bibitem[{\citenamefont{Binder}(1982)}]{binder:1982}
\bibinfo{author}{\bibfnamefont{K.}~\bibnamefont{Binder}},
  \bibinfo{journal}{Phys. Rev. A} \textbf{\bibinfo{volume}{25}},
  \bibinfo{pages}{1699} (\bibinfo{year}{1982}),
  \urlprefix\url{http://dx.doi.org/10.1103/PhysRevA.25.1699}.

\bibitem[{\citenamefont{Lee and Kosterlitz}(1990)}]{physrevlett.65.137}
\bibinfo{author}{\bibfnamefont{J.}~\bibnamefont{Lee}} \bibnamefont{and}
  \bibinfo{author}{\bibfnamefont{J.~M.} \bibnamefont{Kosterlitz}},
  \bibinfo{journal}{Phys. Rev. Lett.} \textbf{\bibinfo{volume}{65}},
  \bibinfo{pages}{137} (\bibinfo{year}{1990}),
  \urlprefix\url{http://dx.doi.org/10.1103/PhysRevLett.65.137}.

\bibitem[{\citenamefont{Kim}(2005)}]{kim:2005}
\bibinfo{author}{\bibfnamefont{Y.~C.} \bibnamefont{Kim}},
  \bibinfo{journal}{Phys.~Rev. E} \textbf{\bibinfo{volume}{71}},
  \bibinfo{pages}{051501} (\bibinfo{year}{2005}), \eprint{cond-mat/0503480},
  \urlprefix\url{http://arxiv.org/abs/cond-mat/0503480}.

\bibitem[{\citenamefont{Domany et~al.}(1984)\citenamefont{Domany, Schick, and
  Swendsen}}]{physrevlett.52.1535}
\bibinfo{author}{\bibfnamefont{E.}~\bibnamefont{Domany}},
  \bibinfo{author}{\bibfnamefont{M.}~\bibnamefont{Schick}}, \bibnamefont{and}
  \bibinfo{author}{\bibfnamefont{R.~H.} \bibnamefont{Swendsen}},
  \bibinfo{journal}{Phys. Rev. Lett.} \textbf{\bibinfo{volume}{52}},
  \bibinfo{pages}{1535} (\bibinfo{year}{1984}),
  \urlprefix\url{http://dx.doi.org/10.1103/PhysRevLett.52.1535}.

\bibitem[{\citenamefont{Jonsson et~al.}(1993)\citenamefont{Jonsson, Minnhagen,
  and Nyl\'{e}n}}]{physrevlett.70.1327}
\bibinfo{author}{\bibfnamefont{A.}~\bibnamefont{Jonsson}},
  \bibinfo{author}{\bibfnamefont{P.}~\bibnamefont{Minnhagen}},
  \bibnamefont{and}
  \bibinfo{author}{\bibfnamefont{M.}~\bibnamefont{Nyl\'{e}n}},
  \bibinfo{journal}{Phys. Rev. Lett.} \textbf{\bibinfo{volume}{70}},
  \bibinfo{pages}{1327} (\bibinfo{year}{1993}),
  \urlprefix\url{http://dx.doi.org/10.1103/PhysRevLett.70.1327}.

\bibitem[{fn2()}]{fn2}
\bibinfo{note}{A. van Enter, private communication (2014)}.

\bibitem[{\citenamefont{Bl\"{o}te et~al.}(2002)\citenamefont{Bl\"{o}te, Guo,
  and Hilhorst}}]{physrevlett.88.047203}
\bibinfo{author}{\bibfnamefont{H.~W.~J.} \bibnamefont{Bl\"{o}te}},
  \bibinfo{author}{\bibfnamefont{W.}~\bibnamefont{Guo}}, \bibnamefont{and}
  \bibinfo{author}{\bibfnamefont{H.~J.} \bibnamefont{Hilhorst}},
  \bibinfo{journal}{Phys. Rev. Lett.} \textbf{\bibinfo{volume}{88}},
  \bibinfo{pages}{047203} (\bibinfo{year}{2002}),
  \urlprefix\url{http://dx.doi.org/10.1103/PhysRevLett.88.047203}.

\bibitem[{\citenamefont{Vink}(2014)}]{Vink2014}
\bibinfo{author}{\bibfnamefont{R.~L.~C.} \bibnamefont{Vink}},
  \bibinfo{journal}{J.~Chem.~Phys.} \textbf{\bibinfo{volume}{140}},
  \bibinfo{pages}{104509} (\bibinfo{year}{2014}), ISSN
  \bibinfo{issn}{1089-7690},
  \urlprefix\url{http://dx.doi.org/10.1063/1.4867897}.

\bibitem[{\citenamefont{Ferrenberg and
  Swendsen}(1989)}]{ferrenberg.swendsen:1989}
\bibinfo{author}{\bibfnamefont{A.~M.} \bibnamefont{Ferrenberg}}
  \bibnamefont{and} \bibinfo{author}{\bibfnamefont{R.~H.}
  \bibnamefont{Swendsen}}, \bibinfo{journal}{Phys.~Rev. Lett.}
  \textbf{\bibinfo{volume}{63}}, \bibinfo{pages}{1195} (\bibinfo{year}{1989}),
  \urlprefix\url{http://dx.doi.org/10.1103/physrevlett.63.1195}.

\end{thebibliography}

\appendix

\section{Kim-Fisher scaling algorithm}

We still describe the Kim-Fisher scaling algorithm~\cite{kim.fisher:2004} that 
was used to generate the data of \fig{fig2}(b,c). For a fixed sharpness 
parameter $p$ and system size $L$, it is straightforward to measure $U_4$ and 
$\rho=\mean{N}/L^2$ as a function of $\mu$. A plot of $U_4$ versus $\rho$, which 
is thus parameterized by $\mu$, reveals two minima. The location of the minimum 
at low density is denoted $\rho^-(L,p)$, with $Q^-(L,p)$ the corresponding 
cumulant value. Similarly, the location of the minimum at high density is 
denoted $\rho^+(L,p)$, with $Q^+(L,p)$ the corresponding cumulant value. The 
purpose of the scaling algorithm is to evaluate the order parameter $\Delta$ as 
a function of $p$ in the thermodynamic limit: $\Delta(p) = \lim_{L \to \infty} 
(\rho^+(L,p) - \rho^-(L,p))/2$. To this end, one defines the quantities
\begin{eqnarray}
Q_{\rm min}(L,p) &\equiv&
  \frac{Q^+(L,p) + Q^-(L,p)}{2} \quad, \\
x(L,p) &\equiv& Q_{\rm min}(L,p)
  \ln \left[ \frac{4}{e Q_{\rm min}(L,p)} \right] \quad, \\
\label{eq:yy} y(L,p) &\equiv&
  \frac{\rho^+(L,p) - \rho^-(L,p)}{\Delta(p)} \quad.
\end{eqnarray}
The algorithm starts in the first-order regime, i.e.~with a large value of $p$. 
The peaks in $P(N)$ are then well separated and the free energy barrier $\Delta 
F $ will be large, as in \fig{fig1}(a). In this regime, it can be shown 
rigorously that the points $(x,y)$ of different system sizes $L$, should all 
collapse onto the line $y = 1+x/2$. Recall that $\Delta(p)$ in \eq{eq:yy} is the 
order parameter in the thermodynamic limit at the considered $p$, precisely the 
quantity of interest, which may thus be obtained by fitting until the best 
collapse onto $y=1+x/2$ occurs. Next, $p$ is chosen closer to the critical 
point, the points $(x,y)$ are calculated as before, but this time around 
$\Delta(p)$ is chosen such that the new data set joins smoothly with the 
previous one, yielding an estimate of the order parameter at the new $p$. This 
procedure is repeated as closely as possible to the tricritical point, where 
$\Delta$ vanishes, yielding an estimate of $\pc$.

\end{document}